\documentclass[11pt,twoside]{article}

\usepackage{galev06}
\usepackage{epsf}
\usepackage{psfig}
\usepackage{lscape}

\begin{document}
\title{Intergalactic Star Formation}   
\author{Pierre-Alain Duc$^{1}$, M\'ed\'eric Boquien$^{1}$, Jonathan Braine$^{2}$, Elias Brinks$^{3}$, Ute Lisenfeld$^{4}$  and Vassilis Charmandaris$^5$}
\affil{$^{1}$AIM  - UMR n$^\circ$ 7158, Service d'astrophysique, CEA--Saclay, France\\
$^{2}$ Observatoire de Bordeaux, UMR n$^\circ$ 5804,  Floirac, France\\
$^{3}$ Centre for Astrophysics Research, University of Hertfordshire, UK\\
$^{4}$ Dept. de F\'\i sica Te\'orica y del Cosmos, U. Granada, Granada, Spain\\
$^{5}$ Department of Physics, University of Crete, Heraklion, Greece
}

\begin{abstract} 
Star formation in interacting systems may take place in various locations, from the dust--enshrouded core of Ultraluminous Infrared Galaxies to more unusual places such as the  debris of colliding galaxies expelled into the intergalactic medium. Determining whether star-formation proceeds in the latter environment, far from the parent galaxies, in a similar way as in spiral disks has motivated the multi--wavelength study presented here. We collected VLA/HI,  UV/GALEX, optical H$\alpha$  and MIR/Spitzer images of a few nearby interacting systems chosen for their prominent  "intergalactic" star formation activity. Preliminary results on the spectacular collisional HI ring around NGC 5291 are presented. 
\end{abstract}

\section{Introduction} 

Interacting systems provide a large variety of environments able to host star--forming regions:  from the densest ones --- the central regions where the accumulation and collapse of gas lead    to the onset of intense, dust enshrouded,  circum-nuclear starbursts --- to the most diffuse ones ---  the tidal tails and collisional debris which extend far out from the parent galaxies. In some cases, these external star---forming regions may even appear detached, somehow isolated in the IGM,  and are qualified as  ``intergalactic". The interface region between  the colliding galaxies provides also a particular environment where cloud--cloud shocks may trigger  star formation.

 It could  be expected that the modes of SF and thus the  Initial Mass Function (IMF) or the Star Formation Efficiency (SFE) would change from one type of region to the other. Surprisingly, it was recently claimed that the SFE may not be so different in the dense core of Ultraluminous Infrared Galaxies and in the more quiescent regions  of spiral arms (Gao \& Solomon~2004).  Whether this is also the case for the regime of ``external" or even ``intergalactic" star--formation is still an open question which has sofar not been studied in detail.

Intergalactic SF regions actually appear as particularly interesting laboratories to study the process of star formation. Indeed, on one hand, they share with galactic SF regions their chemical characteristics --- at first order, they have the same ISM, including the molecular gas found in quantity in tidal tails (Braine et al., 2001); on the other hand, they are by nature detached  and are thus simple objects.  Beside, the observed intergalactic star forming regions are often formed within ``pure" HI clouds, with no evidence for the presence of a pre-existing stellar component from the parent galaxies. Without such  contamination,  the star--formation history can be more easily reconstructed than in galactic disks where several generations of stars coexist. In particular, starburst ages can be  derived with reduced uncertainties. Furthermore, for those SF episodes triggered by a tidal interaction, numerical simulations may provide dynamical ages for the system and thus upper limits on the onset of SF. 

Star Formation may be studied using several indicators: ultraviolet emission, which is sensitive to the SF episodes at time scales of about 100 Myr, H$\alpha$ emission, probing SF over time scales at less than 10 Myr and  mid--infrared one, with time scales probably slightly higher. The far--infrared regime would provide  more direct estimates of the Star Formation Rate, but data in this wavelength domain and for this type of objects will be rare until the launch of Herschel. Comparing the SFRs determined from these different indicators, one may learn about the level of dust obscuration, and, in addition set constraints on the star formation history. Combining these multiwavelength data to construct the Spectral Energy Distribution and comparing it with evolutionary synthesis models, one may even constrain the IMF.

We applied such a method to several interacting systems 
for which we obtained either PI or archival VLA/HI, GALEX UV, Spitzer mid--IR  or 
ground based H$\alpha$ data. The first  more systematic studies of colliding galaxies with Spitzer indicate that globally the level of star--formation along tidal tails is low, contributing  less than 10\% to the total SFR (see contribution by C. Struck in this conference).  In contrast, the objects we have selected exhibit in their surroundings tidal or collisional debris which  is particularly active in forming stars and capable of forming enough of them to build objects as massive as dwarf galaxies (the so-called Tidal Dwarf Galaxies, see Duc et al., 2004, and references therein).
We present here our first analysis of the spectacular system NGC 5291.


\section{A case study: NGC 5291}
NGC 5291 is remarkable for its  prominent HI ring--like structure hosting numerous intergalactic 
HII regions. The ring has a diameter of 180 kpc and contains an HI mass of $6\times10^{10}$ M$_\textrm{\sun}$. The system, which lies in the outskirts of the Abell Cluster 3574, is composed of an early--type galaxy, NGC 5291,  and a highly disturbed companion galaxy, ``the Seashell''.  It is however yet unclear whether the latter object is responsible for the bulls-eye collision at the origin of the ring. Several other nearby cluster  members could account for it. Duc \& Mirabel (1998)  found that the HII regions associated with the HI clumps along the ring had moderate metallicity ($12+\log O/H=8.4-8.6$) and could thus exclude a primordial origin for the gas.

\begin{figure}[t]
\plotone{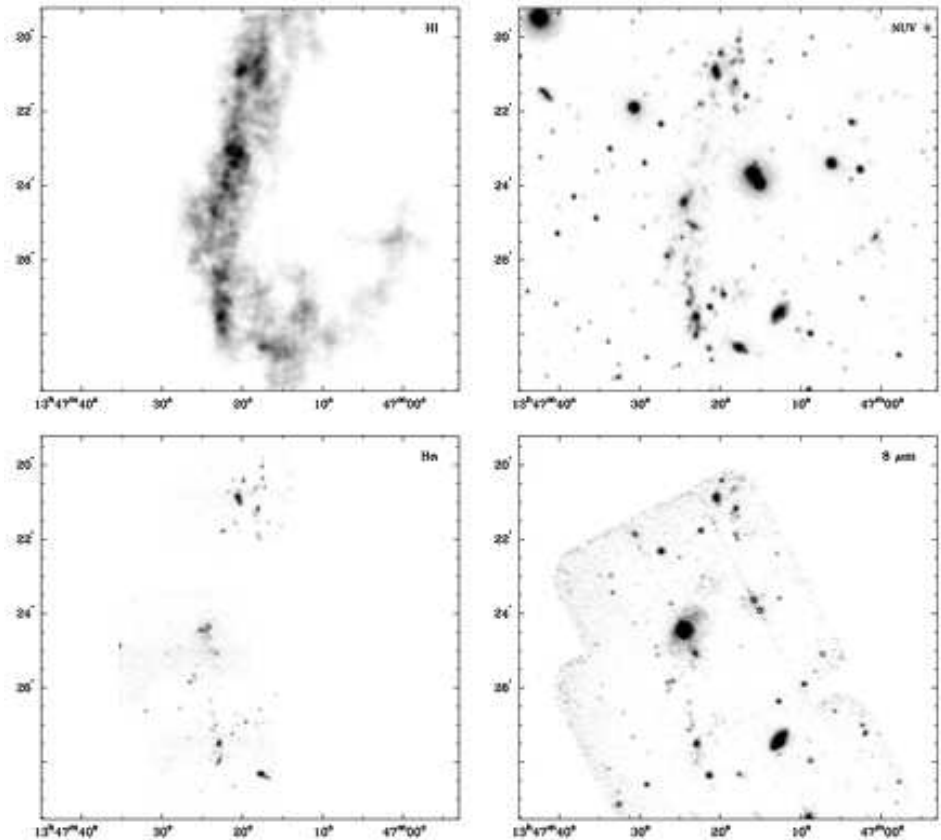}
\caption{The collisional ring around NGC 5291 and associated star--forming regions: VLA HI (top left), GALEX near-ultraviolet (top right), Fabry--Perot H$\alpha$ (bottom left), and Spitzer 8.0 $\mu$m (bottom right).}
\end{figure}

We have obtained high--resolution VLA/HI and GALEX near ultra-violet (NUV) images of the system  which we combined with our previously acquired optical, H$\alpha$ (Bournaud et al. 2004) and archival IRAC/Spitzer images (first published in Higdon et al., 2006).

We identified along the ring 29 HII--regions, most of them lying within dense HI condensations (see Figure 1 and Boquien et al., 2006). We measured  their fluxes  in the NUV, H$\alpha$ and IRAC/8--$\mu$m bands and  compared those different star formation tracers. Their morphology   is remarkably similar, indicating that all of them are fair tracers of star formation. The intergalactic IR emission at 8.0 $\mu$m, dominated by PAH bands (Higdon et al., 2006), when normalized to the hot dust continuum, is comparable to the integrated emission of dwarf galaxies of the same metallicity and to the emission of individual HII regions in spirals. This indicates that the 8.0 $\mu$m band emission is probably an estimator of the SFR that is as reliable for collisional debris as it is  for the objects for which it had sofar been calibrated. Whereas the 8.0 $\mu$m is ``normal'', there is a clear excess of near ultraviolet emission compared to individual HII regions in spirals. The $\left[8.0\right]/\left[NUV\right]$ and $\left[H\alpha\right]/\left[NUV\right]$ flux ratios are low although there are some strong variations from one region to the other. The variations of the  $\left[8.0\right]/\left[NUV\right]$ ratio cannot be explained by a metallicity effect since the oxygen abundance was found to be uniform along the entire HI structure. Moreover, it can neither be due to dust, since correcting for the spatial variations of the dust extinction, as estimated from the HI column density,  does not reduce the scatter in the $\left[H\alpha\right]/\left[NUV\right]$ and $\left[8.0\right]/\left[NUV\right]$ flux ratios. Their variations are best explained by age effects. A model of the evolution of $\left[H\alpha\right]/\left[NUV\right]$ with time favors young instantaneous starbursts: the ultraviolet excess  plus the large scatter indicating rapid variations of the $\left[H\alpha\right]/\left[NUV\right]$ ratio are difficult to explain with a constant SFR or moderately decreasing SFR. So far no indication for the presence of an old stellar component has been found. Comparing the individual starburst ages with the dynamical timescale for the formation of the HI ring in which they lie, we concluded that it is possible that SF has set on quasi--simulatenously over the 180--kpc long gaseous structure.

Using the standard Kennicutt calibrations, we estimated from the UV, uncorrected for dust extinction,  a total SFR around NGC~5291 of 1.3 M$_\textrm{\sun}$/year. 
It appears that about 80\% of the star formation in the entire system takes place in the intergalactic HII regions,  a value much higher than in most nearby interacting systems. 

Finally, the star formation history of the intergalactic HII regions of NGC~5291 is such that simple Single Star Population models (coupled with dust models) should  be  enough to fit  their Spectral Energy Distribution. This will allow us to get some constraints on the IMF in the intergalactic environment. 

\section{Impact of the intergalactic star formation} 
Depending  on their SFR and the amount of available gas reservoir, the star--forming regions present in collisional debris will evolve differently. Most likely the  faintest intergalactic HII regions discovered in nearby groups and clusters (e.g. Cortese et al., 2006), will form stars that will be dispersed in the IGM and thus contribute to the population of intracluster stars. The giant HII regions located along tidal tails may form (Super) Star Clusters and be the progenitors of the young globular clusters found around mergers (Schweizer, 2006). Finally, the most active and massive regions, with SFR exceeding 0.1 M$_\textrm{\sun}$/year  and masses above $10^9$ M$_\textrm{\sun}$, will evolve into dwarf--like galaxies which, if they survive long--enough, will become almost undistinguishable from classical satellite galaxies (Bournaud \& Duc, 2006).




\end{document}